# Deep Learning for Causal Inference

Vikas Ramachandra

Stanford University Graduate School of Business

655 Knight Way, Stanford, CA 94305

## Abstract

In this paper, we propose the use of deep learning techniques in econometrics, specifically for causal inference and for estimating individual as well as average treatment effects. The contribution of this paper is twofold:

1. For generalized neighbor matching to estimate individual and average treatment effects, we analyze the use of autoencoders for dimensionality reduction while maintaining the local neighborhood structure among the data points in the embedding space. This deep learning based technique is shown to perform better than simple k nearest neighbor matching for estimating treatment effects, especially when the data points have several features/covariates but reside in a low dimensional manifold in high dimensional space. We also observe better performance than manifold learning methods for neighbor matching.

2. Propensity score matching is one specific and popular way to perform matching in order to estimate average and individual treatment effects. We propose the use of deep neural networks (DNNs) for propensity score matching, and present a network called PropensityNet for this. This is a generalization of the logistic regression technique traditionally used to estimate propensity scores and we show empirically that DNNs perform better than logistic regression at propensity score matching.

Code for both methods will be made available shortly on Github at: https://github.com/vikas84bf

## 1. The problem of causal inference

We consider a setup where there are n units or data points, indexed by i = (1, . . . , n). We postulate the existence of a pair of potential outcomes for each unit, $(Y_i(0), Y_i(1))$ (following the potential outcome or Rubin Causal Model [4]), with the unit-level causal effect defined as the difference in potential outcomes, $T_i = Y_i(1) - Y_i(0)$. Let $W_i \in \{0, 1\}$ be the binary indicator for the treatment, with $W_i = 0$ indicating that unit i received the control treatment, and $W_i = 1$ indicating that unit i received the active treatment. The realized outcome for unit i is the potential outcome corresponding to the treatment received: $Y(obs)_i = Y_i(W_i) = Y_i(0) \text{ if } W_i = 0, Y_i(1) \text{ if } W_i = 1$.

Let $X_i$ be a N-component vector of features, covariates or pretreatment variables, known not to be affected by the treatment. Our data consist of the triple $(Y(obs)_i, W_i, X_i)$, for i = (1, . . . , n), which are regarded as an i.i.d sample drawn from a large population. We assume that observations are exchangeable, and that there is no interference (the stable unit treatment value assumption, or sutva).

Since we cannot observe the counterfactual for any particular $x_i$ unit, one way to estimate the treatment effect for each unit will be by using values from its neighbors which received the opposite treatment, and by taking the difference between the two outcomes. This individual treatment effect ITE can be written as:

$ITE =$
$T(estimated)_i = Y_i(1) - Y_{neighbor}(0), \text{ if } W_i = 1, \text{ and } -(Y_i(0) - Y_{neighbor}(1)), \text{ if } W_i = 0$

There are different techniques to determine the 'neighbors' in the above construct, and we will look at two such methods: 1. Generalized neighbor matching as well as 2. propensity score based matching, and introduce deep learning based models to do both types of matching.

## 2. Neighbor matching to estimate individual and average treatment effects

As discussed above, the missing counterfactual data problem can be addressed (under certain assumptions [2]) by matching each unit which did not receive treatment (W=0) with its 'nearest' unit from the group that received treatment (W=1) for the binary treatment case. There are various techniques which have been used for matching, propensity score based matching[3] as well as generalized neighbor matching [1] (using clustering, spectral clustering and manifold learning methods).

### 2.1 Propensity score matching

One of the most popular techniques for matching is by using propensity scores [2][3], as briefly described below.

In the statistical analysis of observational data, propensity score matching (PSM) is a statistical matching technique that attempts to estimate the effect of a treatment, policy, or other intervention by accounting for the covariates that predict receiving the treatment. PSM attempts to reduce the bias due to confounding variables that could be found in an estimate of the treatment effect obtained from simply comparing outcomes among units that received the treatment versus those that did not. The technique implements the Rubin causal model for observational studies. The possibility of bias arises because the apparent difference in outcome between these two groups of units may depend on characteristics that affected whether or not a unit received a given treatment instead of due to the effect of the treatment per se. In randomized experiments, the randomization enables unbiased estimation of treatment effects; for each covariate, randomization implies that treatment-groups will be balanced on average, by the law of large numbers. Unfortunately, for observational studies, the assignment of treatments to research subjects is typically not random. Matching attempts to mimic randomization by creating a sample of units that received the treatment that is comparable on all observed covariates to a sample of units that did not receive the treatment, and these two matched groups can be used to estimate the average or individual treatment effect (by taking a difference between the outcomes of the two matched groups or units.)

PSM is for cases of causal inference and simple selection bias in non-experimental settings in which: (i) few units in the non-treatment comparison group are comparable to the treatment units; and (ii) selecting a subset of comparison units similar to the treatment unit is difficult because units must be compared across a high-dimensional set of pretreatment characteristics.

PSM employs a predicted probability of group membership e.g., treatment vs. control group—based on observed predictors, usually obtained from logistic regression to create a counterfactual group.

Traditional procedure for Propensity score matching is as follows:

1. Run logistic regression:

Dependent variable: $Y = 1$, if participate; $Y = 0$, otherwise.
Choose appropriate confounders (variables hypothesized to be associated with both treatment and outcome)
Obtain propensity score: predicted probability (p) or $\log[p/(1 − p)]$.

2. Check that propensity score is balanced across treatment and comparison groups, and check that covariates are balanced across treatment and comparison groups within strata of the propensity score.

3. Match each participant to one or more nonparticipants on propensity score: Traditionally, nearest neighbor matching is used.

### 2.2 Generalized neighbor matching:

It has been shown that with increasing dimensions, propensity score based nearest neighbor matching has increasing bias [2]. To overcome this problem, various alternatives have been proposed in the literature to propensity score matching, such as using random projections [1] and spectral clustering and local linear embeddings [6]. These techniques work well when the data points span a lower dimensional manifold in higher dimensional space.

**Our contributions:**

1.In this paper, we use deep learning based autoencoders for generalized neighbor matching, for estimation of treatment effect for each data point.

We compare the error in estimated treatment using our method with k nearest neighbors, as well as manifold learning techniques, for simulated datasets, and verify that autoencoder based dimensionality reduction and neighbor matching gives lesser error and a better low dimensional representation compared to k nearest neighbors as well as manifold learning methods.

2. In the case of using the propensity score based method for matching, we also propose the use of deep neural networks (DNNs) for step 1 above in lieu of traditional logistic regression, for propensity score estimation, and we present results for simulated datasets to verify the superior performance of the proposed DNN, PropensityNet for this task.

## 3. Autoencoders for generalized neighbor matching

An autoencoder is an artificial neural network used for unsupervised learning of efficient codings of the input data [7]. The aim of an autoencoder is to learn a representation (encoding) for a set of data, typically for the purpose of dimensionality reduction.

### 3.1 Deep learning based clustering: Autoencoders

Architecturally, the simplest form of an autoencoder is a feedforward, non-recurrent neural network very similar to the multilayer perceptron (MLP) – having an input layer, an output layer and one or more hidden layers connecting them – but with the output layer having the same number of nodes as the input layer, and with the purpose of reconstructing its own inputs (instead of predicting the target value). Therefore, autoencoders are unsupervised learning models.

An autoencoder always consists of two parts, the encoder and the decoder, which can be defined as transitions, $(\phi, \psi)$ such that:

$$\phi : X \rightarrow F \text{, encoder}$$
$$\psi : F \rightarrow X \text{, decoder}$$
$$(\phi, \psi) : \text{argmin}_{(\psi,\phi)} \| X - (\psi * \phi)X \|, \text{ in the L-2 norm sense}$$

The nonlinear functional mappings for the encoder and decoder are learnt to minimize the reconstruction error above. The learnt mapping, if it maps the input to a lower dimensional encoding, becomes a form of non-linear dimensionality reduction technique.

The training algorithm for an autoencoder can be summarized as
For each input x,
Do a feed-forward pass to compute activations at all hidden layers, then at the output layer to obtain an output x'
Measure the deviation of x' from the input x (typically using squared error),
Backpropagate the error through the net and perform weight updates.
Repeat the above steps for several epochs until the error reaches below a certain threshold or converges.

## 3.2 Autoencoders for neighbor matching

We build an autoencoder with the following structure. If the input data as N dimensions, the first and last layers of the autoencoder have N neurons, Our aim is to reduce the dimensions, to M, so the middle layers of the autoencoder has M neurons, as shown in the figure below (left).
In the case of our simulated dataset, we have N=3, M=2, and number of data points =1500. This means that the encoder weights will be 2x1500 and the decoder weights will be 1500x2, since the hidden dimension M=2. The training process will try and learn the weights in an iterative fashion, using mean squared error loss function gradient backpropagation.

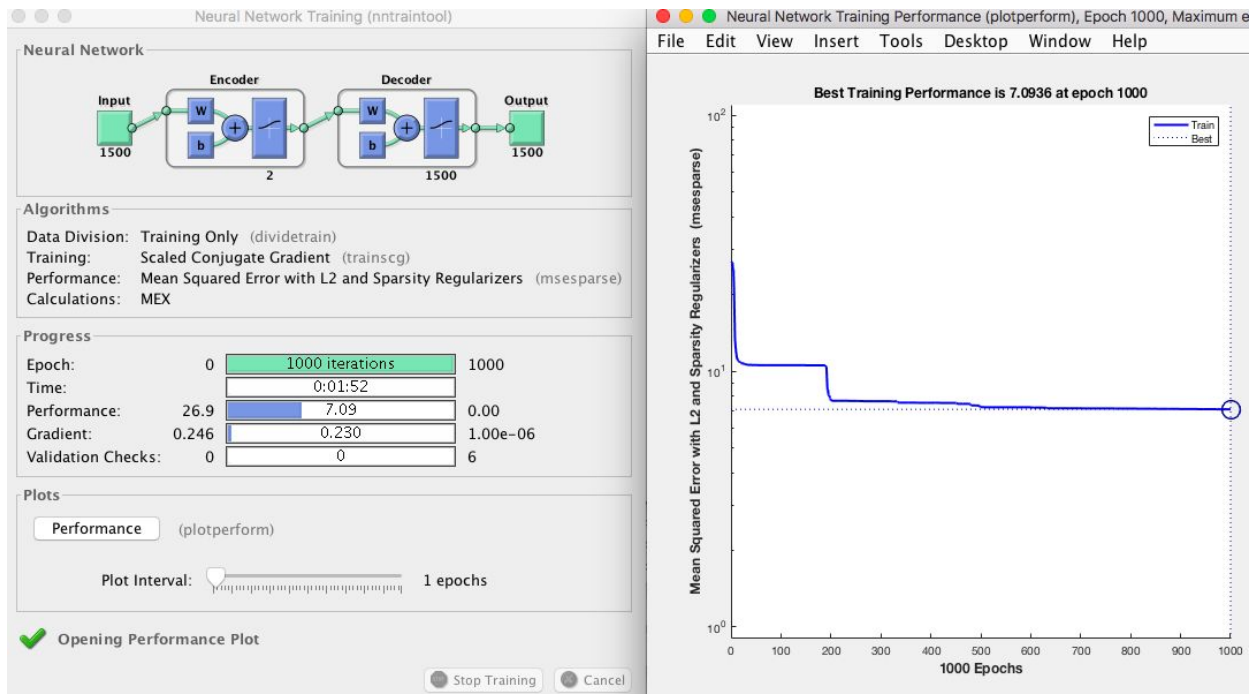

Figure: Left: The autoencoder network, Right: the training mean squared error at each epoch.

In the M dimensional space, individual treatment effect (ITE) is calculated as the difference in the outcomes of the present unit (if treated) and its untreated neighbor(s) in that space, using Euclidean distance in M dim. space to identify neighbors.

ITE: (Y_unit_treated - Y_neighbor__M_dim_untreated).
The above expression for ITE is similar for manifold learning techniques, the main difference being the way we get the mapping to the reduced M dimensional space using manifold learning versus autoencoder techniques.

### 3.3 Experiments and results for generalized neighbor matching

We simulate a dataset as follows. 1500 points are generated, in 3 dimensions. A swiss roll function is applied so that the points lie along a 2D manifold in 2D space, as shown in the figure below.
The generating function f(x) for the Swiss roll is:
$n = 1500;\ t = 3 * \pi/2 * [1 + 2 * rand(n)]\ ;\ h = 11 * rand(n);$
$f(x) = [t * \cos(t), h, t * \sin(t)] + noise$
The data is also split into 6 groups based on the distances from neighbors along manifolds, as shown in the figure below.
For each data point, we assign a binary treatment variable W=0 or 1, & also outcome Y values as a simple linear combination of the x covariates, using 2 different functions based on W=0 or 1.
Then, we project the dataset onto lower dimensions (M=2) using A. autoencoders and B. Manifold learning.

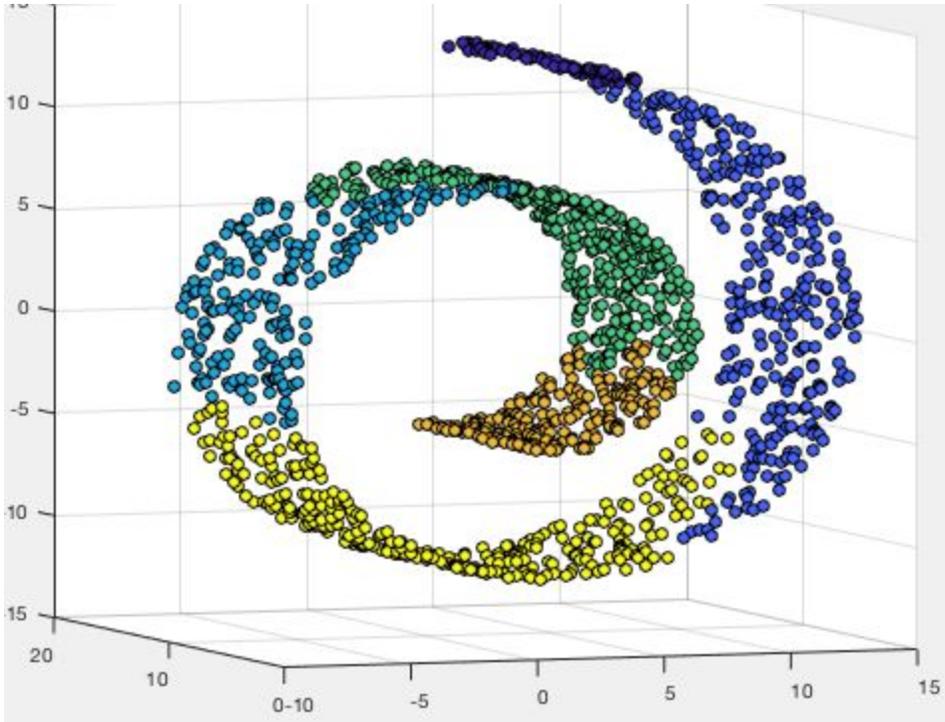

Figure: Original Swiss roll dataset in 3 dimensions used for simulations. Colors show the 6 classes the data was split into for the simulation.

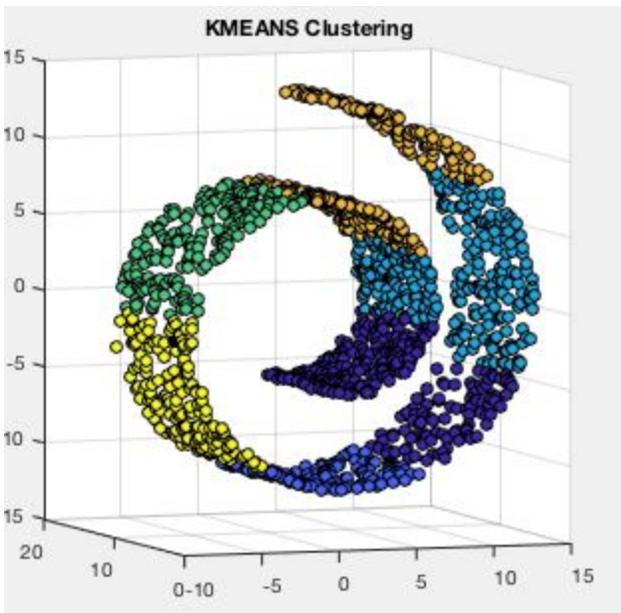

Figure: K-means clustering in the original space for the Swiss roll: it can be seen easily that the algorithm does not learn the manifold nature of the data, and puts far off manifold points from different classes into the same group (color): e.g. the sky blue points. Similarly, k nearest neighbors performs poorly because it does not learn the structure of the data manifold.

When we project the data to 2D space, we visualize the projections.

The figure below shows the dimensionality reduction using A. Principal component analysis (PCA), B. manifold learning (center) based on matrix factorization, and C. Autoencoder.

It is clear that both B. and C. do a good job at learning the structure of the data, unlike PCA, thus a k nearest neighbors in reduced space using Euclidean distance (similar to a PCA decomposition) performs poorly, as it did in the original dimensions.

Next, to compare B. manifold learning and C. autoencoders, We also compute the estimated treatment effect for each point (ITE), and the average absolute error of ITE for B. manifold learning and C. Autoencoder, over all the data points in the test set.

Mean Absolute error (ITE,autoencoder: 3.7127,

Mean absolute error (ITE, Manifold learning): 4.4540

Thus, autoencoder error is 20.27% lesser than manifold learning estimate for the ITE.

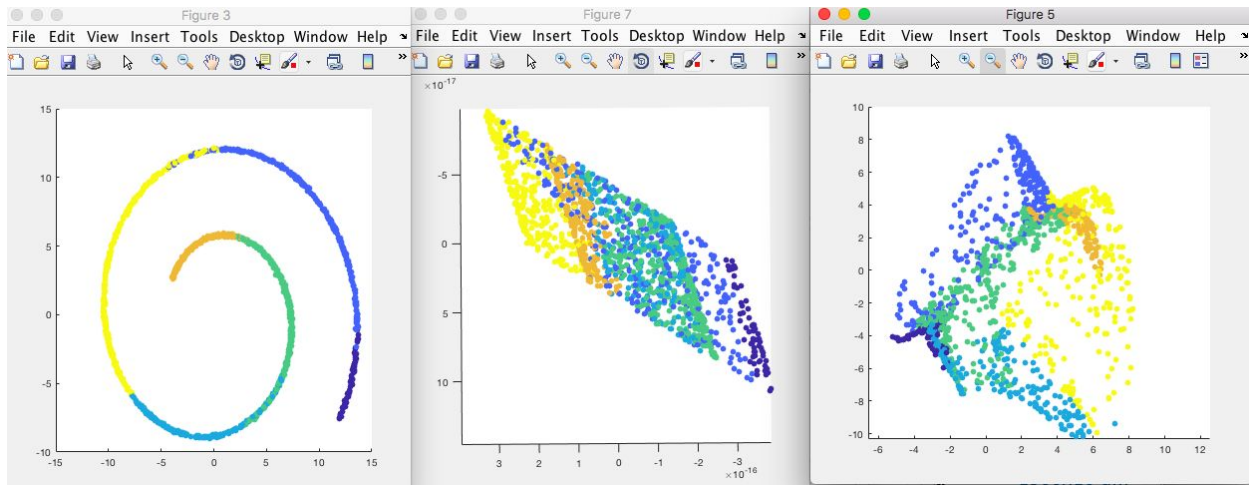

Figure: Clustering and dimensionality reduction using various methods (Same color implies same original assigned group in simulated data)

Left: PCA Center: Manifold learning Right: Autoencoders.

The output from autoencoder also gives the least error in the estimated treatment effect across all units.

## 4. Deep neural networks (DNNs) for propensity score matching

In the above section, we showed how autoencoders for generalized neighbor matching. In this section, we show how deep neural networks for classification can be leveraged to do propensity score matching, specifically to replace logistic regression described in section 2.1

## 4.1 Deep neural networks for classification

A deep neural network (DNN) is an artificial neural network (ANN) with multiple hidden layers between the input and output layers. DNNs can model complex non-linear relationships and can be used for both classification and regression tasks [8]. DNN architectures generate compositional models where the object is expressed as a layered composition of primitives. The extra layers enable composition of features from lower layers, potentially modeling complex data with fewer units than a similarly performing shallow networks or models. DNNs are feedforward networks in which data flows from the input layer to the output layer without looping back. For classification, the last layer of the network is a 'softmax' layer , which outputs the probability of each class. The intermediate layers can be of any form, and the output of each layer is typically passed through a non-linear function.

We can learn the parameters of the classification DNN by using a labeled training dataset, which each data point or unit has a ground truth label. A cost function is specified (such as misclassification error), and the error is back-propagated through the network, to update the weights along the gradient directions iteratively, until we achieve a low error. The learning typically happens in steps for batches of the data (stochastic gradient descent).

The figure below shows an example of the general DNN.

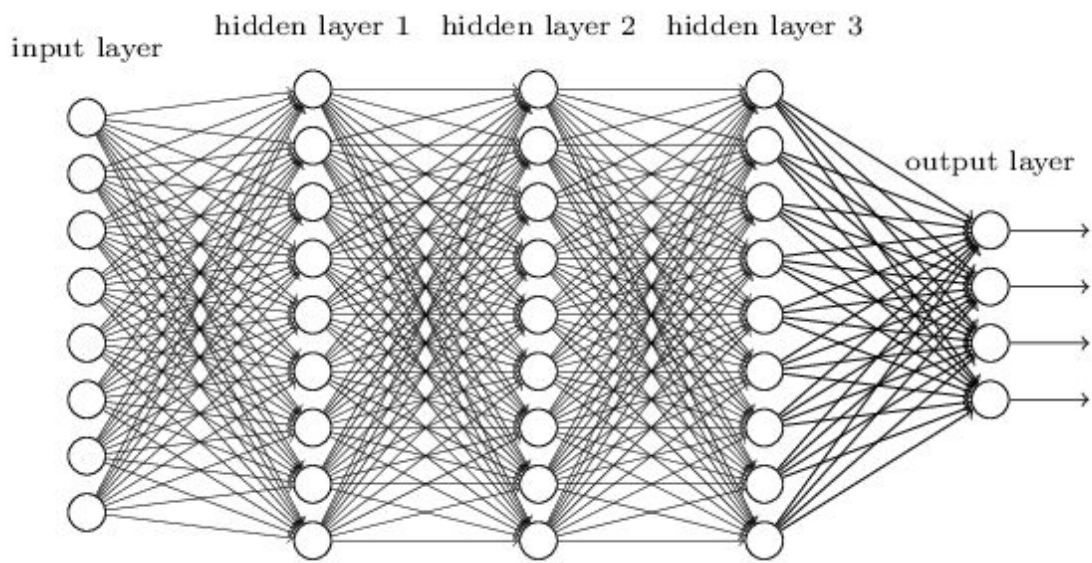

Figure: A general fully connected DNN, for classification.

## 4.2 PropensityNet: Experiments and results for DNN based propensity score matching

We build a DNN 'PropensityNet' to estimate the propensity score, with the inputs being the covariates X as well as the outcome Y across all units. The data is split into training and cross validation folds and categorical cross entropy is used as an error metric (it gives a measure of label misclassification).

We use adadelta as the optimizer algorithm. This DNN PropensityNet is trying to solve a binary classification problem, since the treatment variable W is binary. The output i.e. the last layer (softmax) of the trained network gives us a probability between 0 and 1 for each new/test unit, which is the propensity score. As such, this can be thought of as a generalization of the logistic regression function.

PropensityNet is a fully connected network similar to the above figure, where every neuron in a given layer is connected to every other neuron in the next layer.

The structure of PropensityNet is given below.

```
Model
_________________________________________________________________
Layer (type)                 Output Shape              Param #   
=================================================================
dense_43 (Dense)             (None, 10)                30        
_________________________________________________________________
dropout_31 (Dropout)         (None, 10)                0         
_________________________________________________________________
dense_44 (Dense)             (None, 10)                110       
_________________________________________________________________
dropout_32 (Dropout)         (None, 10)                0         
_________________________________________________________________
dense_45 (Dense)             (None, 10)                110       
_________________________________________________________________
dropout_33 (Dropout)         (None, 10)                0         
_________________________________________________________________
dense_46 (Dense)             (None, 10)                110       
_________________________________________________________________
dropout_34 (Dropout)         (None, 10)                0         
_________________________________________________________________
dense_47 (Dense)             (None, 2)                 22        
=================================================================
Total params: 382
Trainable params: 382
Non-trainable params: 0
_________________________________________________________________
```

Figure: PropensityNet deep neural network model structure

As can be seen above, PropensityNet has 5 dense (fully connected) layers. Each layers also has a dropout of 30%, which is a way to avoid overfitting for DNNs. The output layer is a softmax layer, and gives probability of being in one of the 2 classes (treatment W=1 or 0), which is the propensity score. We have a total of 382 parameters to be trained in the network. The model was built using Keras with Tensorflow backend in R.

We build a simulated dataset as follows.

1000 data points/ units were simulated, with 2 covariates dawn from a uniform distribution, the outcome Y was also randomly drawn from a uniform distribution, and all units assigned to treatment W=1. Thus, we know the ground truth nearest point/neighbor from W=1 for each point in W=0. The unit covariates and outcomes were jittered to get another 1000 units, which were assigned treatment W=0. A logistic regression/logit model was built using W~ X+Y, and the PropensityNet was also trained with W as output, and (X,Y) as inputs for each unit. For both models, we then calculate the assignment error (How far is the test unit assigned on an average from its ground truth neighbor, as well as number of mis-assignments based on estimated propensity score). PropensityNet gave a smaller number of mis-assignments (6% better) as well as a smaller mean absolute misassignment error (12% better, as a percent of ground truth true index of each unit), as well as better accuracy (8% better), compared to logistic regression model, as shown below.

| Model | Mean absolute misclassification error(%) | Number of mis-assignments (%) | Accuracy(%) |
|---|---|---|---|
| Logistic regression | 26.6 | 38 | 62 |
| PropensityNet (DNN) | 19.2 | 26 | 74 |

Table: Various error metrics used to compare the proposed PropensityNet with traditional logit

In the figure below, we plot the control and treatment units based on PropensityNet, to confirm its good performance visually.

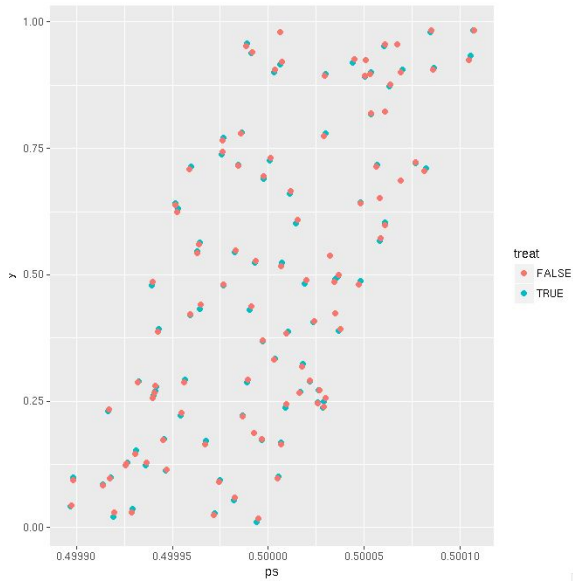

Figure: Plot of a subset of control points (pink), matched (using the PropensityNet output scores) with their neighbors which are treated (blue). It is clear visually that the points are matched well. Y-axis is one covariate and X-axis is propensity score.

## 5. Discussion and conclusion

Recently, there have been several efforts to leverage machine learning techniques for causal inference problems, including estimating heterogeneous treatment effects [5], propensity score modeling as well as neighbor matching [1] for individual treatment effects. Our aim is to contribute to this continuing effort, by adding deep learning techniques to the field of causal inference and econometrics in general. In this paper, we have shown how one can use autoencoders for dimensionality reduction and performing neighbor matching in feature space. We have also built a deep neural network classifier PropensityNet to do propensity score based matching to estimate individual and average treatment effects. The accuracy of both algorithms was verified on simulated datasets. Future work would be to run these algorithms on real world datasets, as well as further leveraging newer deep learning models for causal inference and econometrics. Code for both algorithms will be made available shortly on Github at this location: https://github.com/vikas84bf

## Acknowledgement

We would like to thank Prof. Susan Athey and Prof. Guido Imbens at the Stanford University GSB for several illuminating discussions about causal inference, treatment effects and econometrics.

## References

[1] Matching via Dimensionality Reduction for Estimation of Treatment Effects in Digital Marketing Campaigns; Sheng Li, Nikos Vlassis, Jaya Kawale, Yun Fu, 2016.
https://www.ijcai.org/Proceedings/16/Papers/530.pdf

[2] Large sample properties of matching estimators for average treatment effects; Alberto Abadie and Guido W Imbens, 2001

[3] The central role of the propensity score in observational studies for causal effects; Paul R Rosenbaum and Donald B Rubin, 1983.

[4]Estimating causal effects of treatments in randomized and nonrandomized studies; Donald Rubin, 1974

[5] Recursive Partitioning for Heterogeneous Causal Effects; Susan Athey and Guido W. Imbens, 2015

[6] Robust Propensity Score Computation Method based on Machine Learning with Label-corrupted Data; Chen Wang , Suzhen Wang, Fuyan Shi , Zaixiang Wang, 2018.

[7] Reducing the dimensionality of data with neural networks; G. Hinton and R Salakhutdinov, 2006

[8] Imagenet classification with deep convolutional neural networks; A. Krizhevsky, I, Sutskever, G. Hinton, 2012.